\documentclass[aps,pra,twocolumn,groupedaddress,
showpacs]{revtex4}

\usepackage{graphics}
\usepackage{epsfig}
\usepackage{amsmath}

\begin{document}


\title{Quantum dynamics in single spin measurement}


\author{Stefano Mancini}
\email{stefano.mancini@unicam.it}
\affiliation{Dipartimento di Fisica, Universit\`{a} di Camerino,
I-62032 Camerino, Italy.}
\author{David Vitali}
\email{david.vitali@unicam.it}
\affiliation{Dipartimento di Fisica, Universit\`{a} di Camerino,
I-62032 Camerino, Italy.}
\author{H\'ector  Moya-Cessa}
\email{hmmc@inaoep.mx}
\affiliation{INAOE, Coordinacion de Optica,
Apdo. Postal 51 y 216, 72000 Puebla, Pue., Mexico, \\
and ICTP, Trieste, Italy}

\date{\today}

\begin{abstract}
We study the quantum dynamics of a model for the single-spin measurement in
magnetic-resonance force microscopy. We consider an oscillating driven cantilever
coupled with the magnetic moment of the sample.
Then, the cantilever is damped through an
external bath and its readout is provided by a radiation field.
Conditions for reliable measurements will be discussed.
\end{abstract}

\pacs{03.67.-a, 03.65.Ta, 76.60.-k}

\maketitle

\section{Introduction}\label{intro}

Quantum state manipulation has become a fascinating perspective of modern physics.
With concepts developed in atomic and molecular physics the field has been further
stimulated by the possibility of quantum computation. For this purpose a
number of individual two-state quantum systems (qubits) should be manipulated in a
controlled way. Several physical realizations of qubits have been considered in
recent years, but nano-electronic devices appear particularly promising because they
can be embedded in circuit and scaled up to large number of qubits. In addition to
manipulation, single qubit measurement is an important and delicate question. For
instance, a force resolution of attonewtons is required to detect qubits represented by single
spins \cite{rug}. Nowadays, magnetic resonance force microscopy (MRFM) is striving for this
ultimate goal \cite{mam}. The use of cantilevers to detect small forces is at the
heart of force microscopy in all its various forms \cite{bin}. Therefore, it is one
of the most promising techniques to also achieve single spin detection.

Usually, small magnetic samples of unpaired electron spins
are detected using a MRFM technique based on cyclic adiabatic inversion
(CAI) \cite{rug,wago}, and the quantum dynamics of measurement for this
method has been recently discussed in \cite{bermancai,brun}.
However, a new technique called ``oscillating-cantilever-driven
adiabatic reversals'' (OSCAR) has been proposed and implemented in
\cite{rugoscar}, where a sensitivity of about one hundred spins has been demonstrated.
Very recently, this detection protocol has been improved and a sensitivity equivalent to a single electron spin has been
finally achieved \cite{rugoscarnew}. In the OSCAR protocol, the driven cantilever
causes the adiabatic inversion of the paramagnetic moment of the sample.
The backaction of this moment causes a frequency shift of the cantilever,
which is then detected by a fiber-optic interferometer.
A classical description of the spin-cantilever dynamics in the OSCAR technique has been
given in Ref.~\cite{ber}.
The quantum dynamics of the spin-cantilever system has been instead studied in Ref.~\cite{berman},
where the master equation including the decohering effect of a
high-temperature environment has been solved numerically.
The present paper investigates the quantum dynamics of the whole system including
not only the spin and the cantilever, but also the effective measurement device,
i.e., the fiber-optic interferometer,
which is here represented by a quantized radiation mode.
The spin state manifests itself in the probability
distribution of the phase of the field and our treatment
allows us to take into account also the backaction of the measurement
meter (the field mode) on the spin-cantilever system.
Differently from \cite{berman}, our approach is based on the exact analytical solution
of the quantum dynamics which is valid in the
adiabatic limit and for not too large spin-cantilever coupling.
The paper is organized as follows. In Section \ref{mod}
we introduce the model, then in Section \ref{dyn} we solve
the quantum dynamics.  In Section \ref{qm}
we present the main results. Section IV
is for concluding remarks.

\section{The model}\label{mod}

We consider a model where a ferromagnetic particle is mounted on the
cantilever tip. A permanent magnetic field $B_{z}$ points in the $z$ direction,
while a rf field of magnitude $B_{1}$ rotates in the $x-y$ plane in resonance with
the spin precession around $z$ axis at frequency $\omega_{s}$.
A radiation field is then used to monitor the cantilever's position.
We describe the cantilever interacting with a
single electron spin-$1/2$ as a harmonic oscillator of frequency
$\omega_{c}$, and the radiation field as a single electromagnetic mode
of frequency $\omega_{r}$.
Then, the dimensionless Hamiltonian for such a system
(scaled by the factor $\hbar \omega_c$) can be written as
\begin{eqnarray}\label{H0}
   H&=& a^{\dagger}a + \omega_{r}' b^{\dag}b +\omega_{s}' S_{z}
   \nonumber\\
   &+&\epsilon\left[S_{x}\cos\left(\omega_{s} t\right)
   +S_{y}\sin\left(\omega_{s} t\right)\right]
   \nonumber\\
   &-& \sqrt{2}\eta (a+a^{\dag}) S_z - \kappa b^{\dag}b(a+a^{\dag})\,.
\end{eqnarray}
Here $a$ is the ladder operator of the cantilever
\begin{equation}\label{ac}
   a = \sqrt{\frac{k_{c}}{2\hbar \omega_{c}}}\left(Z+i\frac{\omega_c}{k_c}P\right)\,,
\end{equation}
where $P$ and $Z$ are the momentum and
$z$-coordinate operators of the cantilever, and $k_{c}$ is
the cantilever spring constant. Moreover,
$b$, $b^{\dag}$ are the ladder operators of the
radiation mode with $[{b},{b}^{\dagger}]=1$, $\omega_r'
= \omega_r/\omega_c$, $\omega_s'=\omega_s/\omega_c$,
and $S_i$ ($i=x,y,z$) are the usual spin operators
obeying the commutation rule $[S_x,S_y]=2iS_z$ and its cyclic
permutations.
The first row of Eq.~(\ref{H0}) is the free Hamiltonian of the system,
the second row is the driving term of amplitude
$\epsilon$ given by \cite{ber,berman}
\begin{equation}\label{epsilon}
  \epsilon = \frac{\gamma_e B_1}{\omega_c},
\end{equation}
where $\gamma_e$ is the electron gyromagnetic ratio.
Finally, the third row contains the spin-cantilever interaction (due
to the spatially varying magnetic field $B_{z}$) with coupling constant \cite{ber,berman}
\begin{equation}\label{eta}
  \eta = \frac{\gamma_e}{2} \sqrt{\frac{\hbar}{\omega_c k_c}}\left|\frac{\partial B_z}{\partial z}\right|,
\end{equation}
and the optomechanical interaction (due to the radiation pressure
force) with coupling constant $\kappa$, which, for a cavity interferometer
with length $L$ is given by \cite{pond}
\begin{equation}\label{opto}
 \kappa=\frac{\omega_r}{L}\sqrt{\frac{\hbar}{2k_c \omega_c}}.
\end{equation}

The Hamiltonian (\ref{H0})
in interaction picture with respect to $H_0= \omega_{r}' b^{\dag}b +\omega_{s}' S_{z}$ reads
\begin{equation}\label{Hini}
   H= a^{\dag}a +
   \epsilon S_x  - \sqrt{2}\eta (a+a^{\dag}) S_z - \kappa b^{\dag}b(a+a^{\dag})\,,
\end{equation}
which is the same model Hamiltonian considered in \cite{berman}, supplemented with the optomechanical
interaction with the readout radiation mode $b$.
It is then convenient to apply the $\pi/2$ rotation around the $y$ axis in the spin space $
   R=\exp\left[{iS_y(\pi/4)}\right]$ exchanging $S_x$ and $S_z$,
to obtain the transformed Hamiltonian
\begin{eqnarray}\label{Hrot}
   {\widetilde H}&\equiv&RHR^{\dag}\nonumber\\
   &=& a^{\dag}a
   +\epsilon S_z -\left( \sqrt{2}\eta S_x  + \kappa b^{\dag}b\right)(a+a^{\dag})\,.
\end{eqnarray}
Apart from the additional optomechanical interaction term, Eq.~(\ref{Hrot})
is equivalent to the Hamiltonian of a two-level atom with energy separation $2\epsilon$ interacting
with a quantized radiation mode with unit oscillation
frequency, represented by the cantilever $z$-motion \cite{coh}.
The equivalent dipole-interaction term in Eq.~(\ref{Hrot})
contains also counter-rotating terms, which are usually neglected
in quantum optical situations (rotating wave approximation (RWA)) \cite{coh}.
This approximation is however
not justified in the present case. In fact, in typical MRFM situations,
the spin is much faster than the cantilever and adiabatically
follows the cantilever motion, i.e., $\epsilon \gg 1$.
To be more precise, the OSCAR technique is usually operated under the adiabatic condition
$\epsilon^2\gg\eta\alpha_m$, where $\alpha_m$ is the mean amplitude of the cantilever oscillations,
and under the full reversal condition $\epsilon\ll\eta\alpha_m$ \cite{rugoscar,berman}.
However, Ref.~\cite{berman} shows that the best signal sensitivity is achieved for partial
adiabatic reversals, i.e. $ \epsilon \simeq\eta\alpha_m$.
Along this line, we consider the limit $\epsilon\gg1+\eta$ together with $\kappa\ll 1$,
which allow us to treat the interaction part of Eq.~(\ref{Hrot}) as a small perturbation with respect to the free Hamiltonian $a^{\dag}a+\epsilon S_z$. Thus, keeping the lowest nonzero order terms 
(first order in $\kappa$ and second order in $\eta$), 
we end up with the effective Hamiltonian
\begin{eqnarray}
   \widetilde{H}_{eff}=a^{\dagger}a
   +\epsilon S_z+ \chi a^{\dagger}a S_z
   -\kappa b^{\dag}b(a+a^{\dag}),
   \label{Heff}
\end{eqnarray}
where the effective spin--cantilever interaction $\chi$ is given by
\begin{equation}\label{chi}
   \chi = \frac{16\eta^2\epsilon}{4\epsilon^2-1}\,.
\end{equation}
The third term of  Eq.(\ref{Heff}),
describes a dispersive interaction in which the spin and the cantilever
energies are independently conserved and
one induces a state--dependent phase shift on the other.
In particular, the cantilever undergoes a frequency shift $\pm \chi$
depending upon the value of $S_z$, which is just the
frequency shift which is detected in the OSCAR technique by means
of a phase sensitive measurement of the radiation mode.
Therefore we expect that by measuring the probability
distribution of the field phase one effectively measures the spin component $S_z$,
which corresponds to $S_x$ in the initial, non-rotated frame.
This is consistent with the classical description of MRFM \cite{ber},
in which the cantilever always measures the spin component along
the effective magnetic field which, in the strong adiabatic limit of very large
$\epsilon$ considered here, is essentially aligned along the $x$ direction.

\section{Quantum dynamics}\label{dyn}

To describe a realistic situation
we consider the cantilever
plunged in a thermal bath at
equilibrium temperature $T$ and subject to a viscous force with
damping rate $\Gamma$ (with $\Gamma/\omega_{c}\equiv\gamma$).
This situation corresponds to an ohmic environment \cite{CL83}, and since for typical MRFM cantilevers $\omega_c $ is of order of some kHz,
it is $N\equiv k_BT /\hbar \omega_c -1/2 \gg 1$ ($k_{B}$ is the Boltzmann's constant) even at cryogenic temperatures,
so that environmental effects are satisfactorily described
in terms of the Caldeira-Leggett master equation \cite{CL87},
which is valid in the high temperature limit and is described
by the superoperator \cite{qnoise}
\begin{eqnarray}\label{calL}
   {\cal{L}} {\widetilde\rho} &=& -\frac{\gamma}{4}
   \left[a+a^{\dag},\left\{a-a^{\dag},{\widetilde\rho}\right\}\right]
  \nonumber\\
  &&-\frac{\gamma}{2}\frac{k_{B}T}{\hbar\omega_{c}}
  \left[a+a^{\dag},\left[a+a^{\dag},{\widetilde\rho}\right]\right]\,,
\end{eqnarray}
where ${\widetilde\rho}$ is the density matrix of the rotated system.
Therefore, the complete system dynamics will be governed
by the following master equation
\begin{equation}\label{mas}
   \frac{d}{dt} {\widetilde\rho}=-i\left[ {\tilde H}_{eff},
   {\widetilde\rho}\right]+
   {\cal{L}} {\widetilde\rho},
\end{equation}
where, due to the adoption of the dimensionless Hamiltonian of
Eq.~(\ref{Heff}), time $t$ is also dimensionless, i.e., scaled by $\omega_{c}$.

In order to solve the master equation (\ref{mas}) we introduce the
eigenstates $|e\rangle$, $|g\rangle$ of $S_{z}$
with eigenvalues $+1$,~$-1$ respectively,
and we split the total density operator as
\begin{eqnarray}
{\widetilde\rho}&=&{\widetilde\rho}^{\,(ee)}\otimes|e\rangle\langle e|
+{\widetilde\rho}^{\,(gg)}\otimes|g\rangle\langle g|
\nonumber\\
&+&{\widetilde\rho}^{\,(eg)}\otimes|e\rangle\langle g|
+{\widetilde\rho}^{\,(ge)}\otimes|g\rangle\langle e|\,,
\label{split}
\end{eqnarray}
with ${\widetilde\rho}^{\,(ge)}=
\left[{\widetilde\rho}^{\,(eg)}\right]^{\dag}$.

The terms ${\widetilde\rho}^{\,(ij)}$ ($i,j=e,g$)
in Eq.~(\ref{split}) can be expanded
on the set of Fock states external product
operators $|n\rangle_{r}\langle m|$ ($n,m \in {\mathbf N}$)
for the radiation mode and on the continuous set of displacement operators
$D_{c}(\lambda)= \exp\{\lambda a^{\dagger} -\lambda^* a\}$ ($\lambda \in {\bf C}$) for the cantilever mode,
as follow \cite{glau}
\begin{equation}\label{sol}
{\widetilde\rho}^{\,(ij)}(t)=
\sum_{n,m=0}^{\infty}\int\frac{d^{2}\lambda}{\pi}
e^{-|\lambda|^{2}/2}\Theta_{nm}^{\,(ij)}
(\lambda,t) D_{c}^{\dag}(\lambda)
\otimes
|n\rangle_{r}\langle m|\,,
\end{equation}
where $\Theta_{nm}^{(ij)}(t)$ represent
normally ordered characteristic functions \cite{qnoise,glau}
for the cantilever mode for given spin and radiation
states (they are explicitly
given in the Appendix \ref{app}).

From Eq.~(\ref{split}) one gets the non-rotated solution in the original frame
\begin{eqnarray}\label{rhosol}
&&\rho(t)=R^{\dag}{\widetilde\rho}(t)R
\\
&&=\frac{1}{2}\left[
{\widetilde\rho}^{\,(ee)}(t)+{\widetilde\rho}^{\,(gg)}(t)
-{\widetilde\rho}^{\,(eg)}(t)-{\widetilde\rho}^{\,(ge)}(t)
\right]\otimes |e\rangle\langle e|
\nonumber\\
&&+\frac{1}{2}\left[
{\widetilde\rho}^{\,(ee)}(t)+{\widetilde\rho}^{\,(gg)}(t)
+{\widetilde\rho}^{\,(eg)}(t)+{\widetilde\rho}^{\,(ge)}(t)
\right]\otimes |g\rangle\langle g|
\nonumber\\
&&+\frac{1}{2}\left[
{\widetilde\rho}^{\,(ee)}(t)-{\widetilde\rho}^{\,(gg)}(t)
+{\widetilde\rho}^{\,(eg)}(t)-{\widetilde\rho}^{\,(ge)}(t)
\right]\otimes |e\rangle\langle g|
\nonumber\\
&&+\frac{1}{2}\left[
{\widetilde\rho}^{\,(ee)}(t)-{\widetilde\rho}^{\,(gg)}(t)
-{\widetilde\rho}^{\,(eg)}(t)+{\widetilde\rho}^{\,(ge)}(t)
\right]\otimes |g\rangle\langle e|\,.
\nonumber
\end{eqnarray}
As initial condition we consider a coherent state $|\beta\rangle_{r}$
for the radiation field, while for the cantilever we consider a thermal state at temperature $T$,
shifted by a complex amplitude $\alpha$, where ${\rm Re}\{\alpha\}$ gives the initial mean value of the position and ${\rm Im}\{\alpha\}$
the initial mean value of the momentum.

\section{Quantum Measurement}\label{qm}

As discussed above, the cantilever's position carries information about the spin state
and, in turn, it is read by the radiation field. Due to the
optomechanical coupling, the cantilever affects the phase of the
field and we have therefore to look at the probability distribution of the latter.
The canonical probability operator measure (POM)
for phase measurement is \cite{qopt}
\begin{equation}
\Pi(\theta)=\frac{1}{2\pi}|\theta\rangle_{r}\langle\theta|\,,
\quad {\rm with} \quad
|\theta\rangle_{r}=\sum_{n=0}^{\infty} e^{in\theta}|n\rangle_{r}\,.
\end{equation}
Hence, the phase probability distribution results
\begin{equation}\label{Ptheta}
P(\theta,\tau)\equiv{\rm Tr}\left\{\Pi(\theta)\rho(\tau)\right\}
=P_{ee}(\theta,\tau)+P_{gg}(\theta,\tau)\,,
\end{equation}
where we have introduced the new time $\tau\equiv\gamma t$
which, due to the scaling chosen above, is a dimensionless time now scaled by the cantilever damping rate $\Gamma$, and
\begin{eqnarray}
P_{jj}(\theta,\tau)={\rm Tr}\left\{\Pi(\theta){\widetilde\rho}^{\,(jj)}
(\tau)\right\}\,,
\end{eqnarray}
with $j=e,g$.
From Eq.~(\ref{sol}) we get
\begin{eqnarray}
P_{jj}(\theta,\tau)=
\frac{1}{2\pi}\sum_{n,m=0}^{\infty}e^{-i(n-m)\theta}
\,\Theta_{nm}^{\,(jj)}(\lambda=0,\tau)
\,.
\end{eqnarray}

\begin{figure}
\begin{center}
\includegraphics[width=0.45\textwidth]{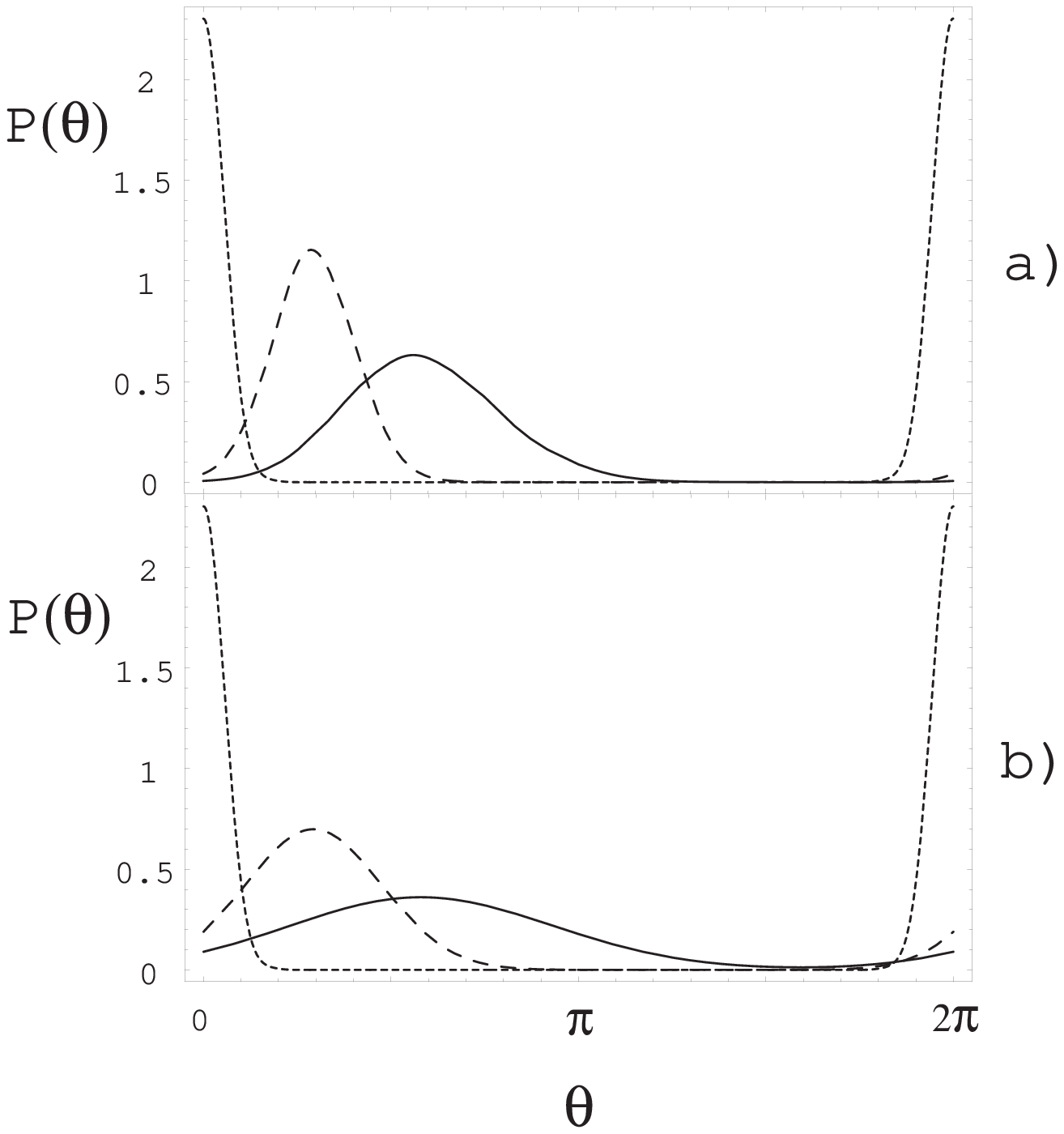}
\end{center}
\caption{\label{fig1} Phase distribution for an initial state of the spin
$(|g\rangle-|e\rangle)/\sqrt{2}$ (in the non-rotated frame) at times
$\tau=0$ (dotted line), $\tau=8\times 10^{4}$ (dashed and solid line).
The dashed line refers to $\chi=0$ while the solid line to $\chi=0.5$.
The values of other parameters are $\gamma=10^{-4}$, $\kappa/\gamma=0.08$,
$\alpha=4i$, $\beta=3$,
and $N=10^{2}$ a), $N=10^{4}$ b).}
\end{figure}

Let us now determine the conditions under which the
apparatus performs reliable measurements
of the spin. As discussed above, the apparatus is expected
to measure $S_x$ (in the original non-rotated frame) and
therefore the easiest case is when the initial state of the
spin is an eigenstate of $S_x$ (in the non-rotated frame).
We consider, for example, the eigenstate with $S_x=-1$, which
correspond to state $|g\rangle $ in the rotated frame and therefore
implying
\begin{eqnarray}\label{iniee}
&&\Theta_{nm}^{\,(jj)}(\lambda,\tau=0)=
e^{-|\beta|^{2}}\frac{\beta^n\left(\beta^{*}\right)^{m}}
{\sqrt{n!m!}}\nonumber\\
&&\times
\exp\left\{
\alpha^{*}\lambda-\alpha^{*}\lambda-\frac{k_{B}T}{\hbar\omega_{c}}
|\lambda|^{2}\right\}\delta_{j,g}\,.
\end{eqnarray}
As a consequence, $S_z=-1$ (in the rotated frame) is a constant of motion and one
has a cantilever frequency shift $\chi$ (see Eq.~(\ref{Heff}))
which manifests itself as a shift of the phase distribution
(see Fig.~\ref{fig1}).
In fact, the phase probability distribution is initially peaked around
$\theta =0$, due to the real amplitude ($\beta =3$) of the initial coherent
state of the radiation mode.
Then the peak moves, toward larger values of $\theta$,
and faster than that with $\chi=0$.
If we had started from state $|e\rangle $ ($S_z=1$ in the rotated frame),
one would have a smaller phase shift of the peak.
This could be surprising as one would expect a greater phase shift
for frequency $(1+\chi)$ and a smaller one for $(1-\chi)$.
However, if one just considers the unitary evolution arising from
Eq.(\ref{Heff}), one reduces to the model studied in
Refs.\cite{pond}, where it is possible to
argue that the induced phase shift is
inversely proportional to the cantilever's frequency.

In Fig.~\ref{fig1} we have considered two different values of temperature,
$N=10^2$ (a) and $N=10^4$ (b), which for typical cantilever frequencies
$\omega_c/2\pi \simeq 10^4$, correspond to $T \simeq 60$ $\mu$K and
$T \simeq 6$ mK respectively.
One sees that already at mK temperatures the peak in the phase
distribution corresponding to the state of the single spin
starts to become too wide and flat to be clearly distinguished.

\begin{figure}
\begin{center}
\includegraphics[width=0.45\textwidth]{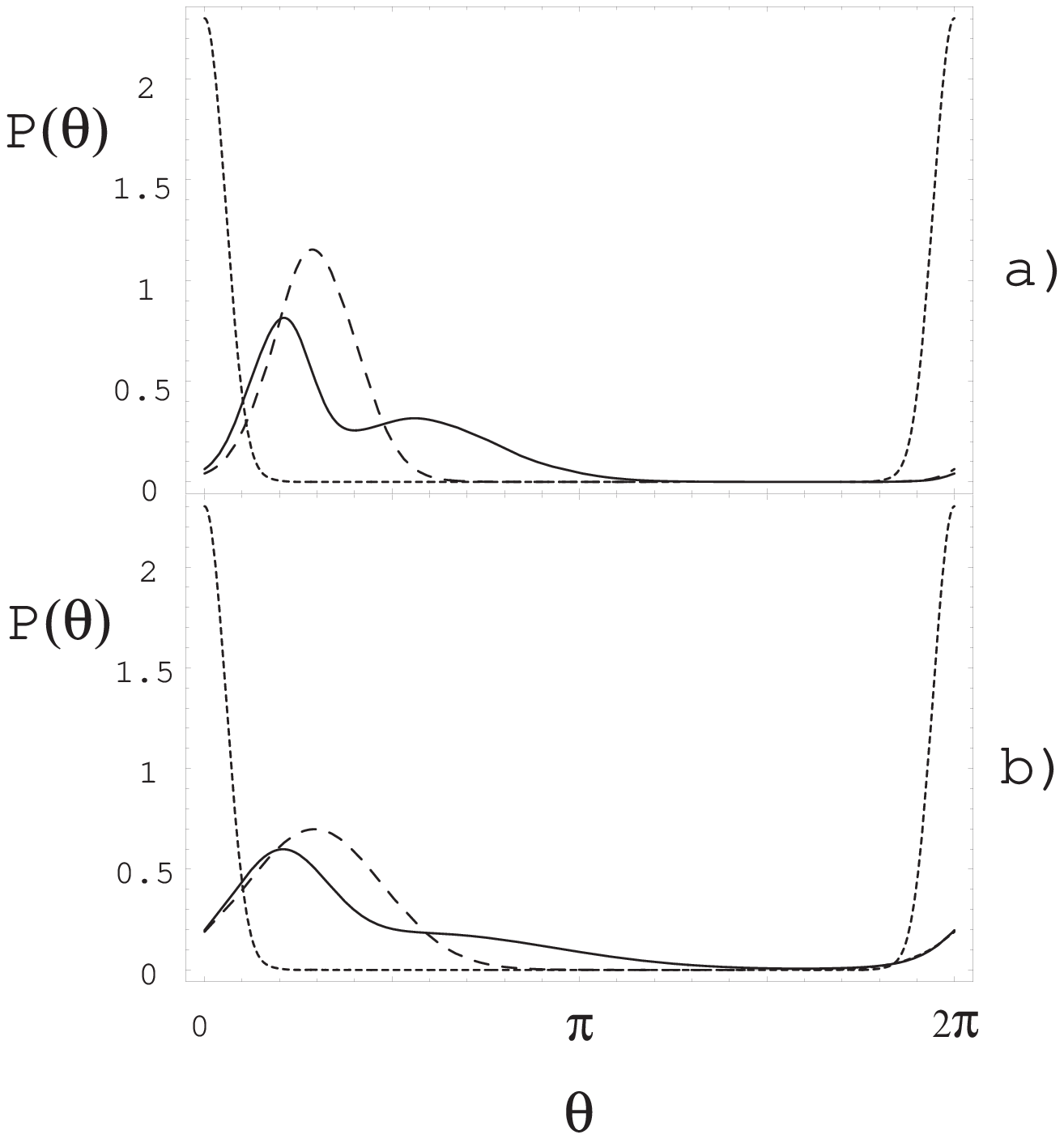}
\end{center}
\caption{\label{fig2} Phase distribution for an initial state of the spin
$|e\rangle$ (in the non-rotated frame) at times
$\tau=0$ (dotted line), $\tau=8\times 10^{4}$ (dashed and solid line).
The dashed line refers to $\chi=0$ while the solid line to $\chi=0.5$.
The values of other parameters are $\gamma=10^{-4}$, $\kappa/\gamma=0.08$,
$\alpha=4i$, $\beta=3$,
and $N=10^{2}$ a), $N=10^{4}$ b).}
\end{figure}

The conditions which the various
parameters have to satisfy in order to have an unambiguous
discrimination of the two spin eigenstates can be better established if we
consider a different initial condition for the spin, i.e.
an initial superposition of the two eigenstates of $S_x$
(in the non-rotated frame). Fig.~\ref{fig2} refers to the initial condition
$S_z=1$ in the initial frame, corresponding to the state
$(|g\rangle+|e\rangle)/\sqrt{2}$ in the rotated frame.
This implies
\begin{eqnarray}\label{inieg}
&&\Theta_{nm}^{\,(jj)}(\lambda,\tau=0)=
e^{-|\beta|^{2}}\frac{\beta^n\left(\beta^{*}\right)^{m}}
{\sqrt{n!m!}}\nonumber\\
&&\times
\exp\left\{
\alpha^{*}\lambda-\alpha^{*}\lambda-\frac{k_{B}T}{\hbar\omega_{c}}
|\lambda|^{2}\right\}\,,
\qquad j=e,g\,.
\end{eqnarray}
Due to the superposition, the measurement performed by the MRFM
is reliable when we
have the simultaneous presence
of two distinct peaks in the phase distribution,
associated with the two different
phase shifts induced by the two spin components. This means
that the measurement time has not to be too small,
in order to allow the two peaks
to become distinguishable. This implies
the condition $\chi t=\chi(\tau/\gamma) > 1$.
Then, reliable measurement can be performed until
the damping affects the dynamics,
as it is confirmed by Fig.~\ref{fig2}a,
where one still has two distinct peaks corresponding to the
two spin states when $\gamma\tau=8$.
In Fig.~\ref{fig2}a the state of the single spin is well detected because
we consider the low temperature condition $N=10^2$,
i.e., $T \simeq 60$ $\mu$K (see above).
At higher temperatures, the effect of the thermal
environment is instead more destructive,
as it can be seen from Fig.~\ref{fig2}b,
which corresponds to $N=10^4$ ($T \simeq 6$ mK), where
the two peaks become too broad and are no more distinguishable.

The other parameter values in Figs.~1 and 2 are $\chi=0.5$
and $\gamma=10^{-4}$. This value of $\gamma$ is realistic,
because $\gamma^{-1}$ corresponds to
the cantilever's quality factor $Q$, which is even higher in recent experiments (see Ref.~\cite{rugoscarnew}), while the chosen value of $\chi$ is still far from
present experimental values. In fact, from Eq.~(\ref{chi}) one has $\chi \simeq 4\eta^2/\epsilon$ and using Eqs.~(\ref{epsilon}) and (\ref{eta}) and the experimental
values of Ref.~\cite{rugoscarnew}, one gets $\epsilon \sim 800$, $\eta \sim 0.04$, so that $\chi \sim 10^{-5}$. Unfortunately, for such a small value of $\chi$,
the radiation field phase shift due to a single spin is too small to be detected and therefore larger values of $\chi$, which means larger values
of the coupling constant $\eta$,
are needed to allow a reliable detection
of the spin state within a reasonable detection time.
Also a larger number of photons in the radiation mode
would improve the resolution
because in such a case the peaks in phase distributions
would be narrower.

\begin{figure}
\begin{center}
\includegraphics[width=0.45\textwidth]{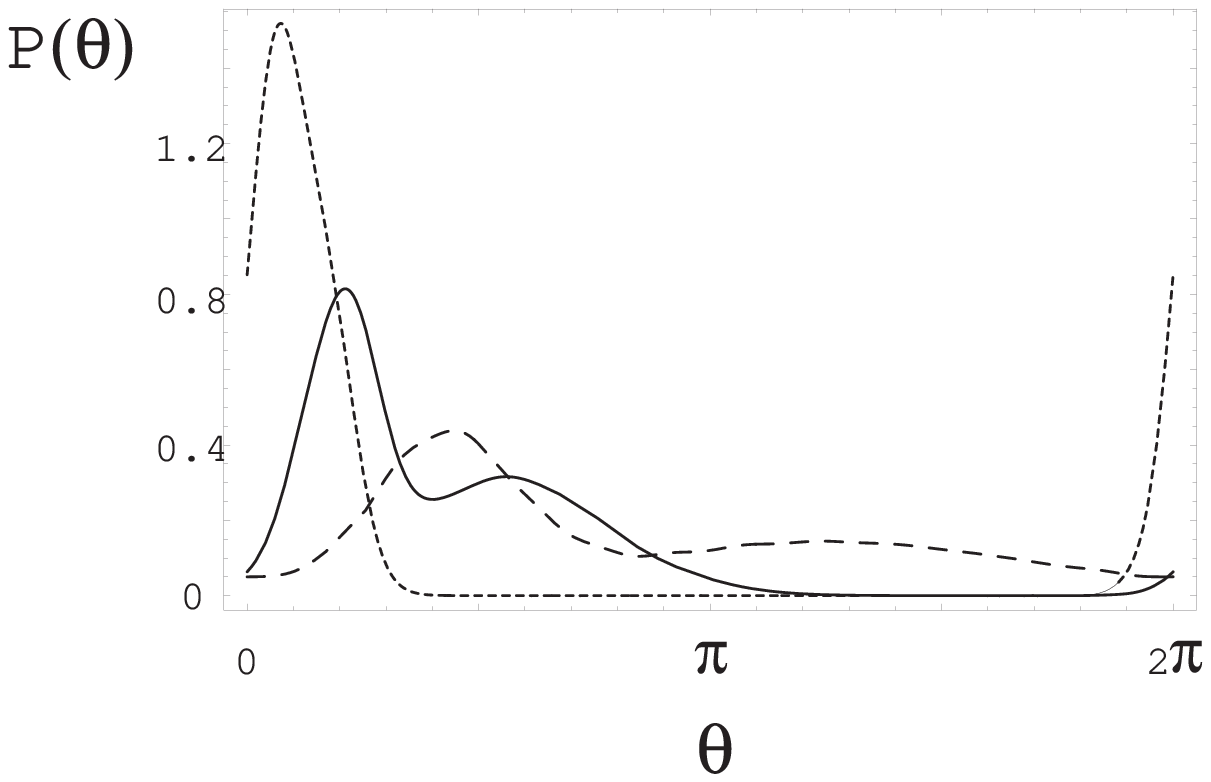}
\end{center}
\caption{\label{fig3} Phase distribution for an initial state of the spin
$|e\rangle$ (in the non-rotated frame)
for different values of $\kappa$ at time $\tau=8\times 10^{4}$.
It is $\kappa/\gamma=0.04$ dotted line, $\kappa/\gamma=0.08$ solid line
and $\kappa/\gamma=0.12$ dashed line.
The values of other parameters are $\gamma=10^{-4}$, $\chi=0.5$,
$\alpha=4i$, $\beta=3$, $N=10^{2}$.}
\end{figure}

Finally, another important parameter is the optomechanical coupling $\kappa$ between
cantilever and the radiation readout mode.
Similarly to $\chi$, this parameter has to be sufficiently large, because otherwise
the phase measurement has not enough sensitivity
(see the dotted line in Fig.~\ref{fig3}, where the two peaks are not well resolved).
On the other hand, the readout mode is a quantum system and
its backaction can blur the measurement result. This happens if the coupling
$\kappa$ is too large, as it is shown by the dashed line of
Fig.~\ref{fig3}. In this case
the backaction is responsible for a phase diffusion of the peaks which are
washed out and no more distinguishable.
By inspection from Eq.(\ref{Heff})
the condition $\kappa |\beta|^2\ll|\alpha|^2$ should
guarantee a negligible optomechanical backaction.
However, it must be modified to take into account
the elapsed time whenever the measurement is performed after many
cycles, which naively implies $\kappa \tau|\beta|^2\ll|\alpha|^2$
(confirmed by the values in Fig.~\ref{fig3}).

\section{Conclusions}

We have presented a model for the description of the quantum dynamics
of the measurement of a single spin by a MRFM operating with
the OSCAR technique \cite{rugoscar}. Our treatment includes also the
phase-sensitive readout of the cantilever frequency-shift
provided by a quantized radiation mode describing the measurement
performed by an optical interferometer.
We have solved analytically the dynamics in the strong adiabatic limit
of a very fast spin, in which the spin reversal cannot be complete, in the presence of a high-temperature ohmic environment acting on the cantilever.
In this limit, the probability distribution of the phase of the optical
readout mode gives a clear signature of the spin component along the effective
magnetic field provided that the effective spin-cantilever coupling
coefficient $\chi$ of Eq.~(\ref{chi}) is sufficiently large
and the optomechanical coupling $\kappa$ is either not too large (in order to limit the backaction)
or not too small (in order to extract information from the system).

\begin{acknowledgments}

H. M.-C. would like to thank the  University of Camerino
for the kind hospitality.

\end{acknowledgments}

\appendix

\section{}\label{app}

Since only $\Theta_{nm}^{(ee)}$
and $\Theta_{nm}^{(gg)}$
enter in Eq.~(\ref{Ptheta}),
we need to solve Eq.~(\ref{mas})
for ${\widetilde\rho}^{\,(ee)}$ and ${\widetilde\rho}^{\,(gg)}$ only.

\subsection{Solution for ${\tilde\rho}^{\,(ee)}$}

Let us define
${\widetilde\rho}^{\,(ee)}_{nm}\equiv{}_{r}\langle n|
{\widetilde\rho}^{\,(ee)}|m\rangle_{r}$, then
Eq.~(\ref{mas}) can be rewritten in the form
\begin{eqnarray}\label{masee}
   \frac{d}{d\tau} {\widetilde\rho}^{\,(ee)}_{nm}
   &=&-i\frac{1+\chi}{\gamma}\left[a^{\dag}a,
   {\widetilde\rho}^{\,(ee)}_{nm}\right]
   \nonumber\\
   &+&i\frac{\kappa}{\gamma} n(a+a^{\dag}){\widetilde\rho}^{\,(ee)}_{nm}
   -i\frac{\kappa}{\gamma} m{\widetilde\rho}^{\,(ee)}_{nm}(a+a^{\dag})
   \nonumber\\
   &-&\frac{1}{4}\left[a^{2}-\left(a^{\dag}\right)^{2},
   {\widetilde\rho}^{\,(ee)}_{nm}\right]
   \nonumber\\
   &+&\frac{1}{2}\left(N+1\right)\left(
   2a{\widetilde\rho}^{\,(ee)}_{nm}a^{\dag}
   -a^{\dag}a{\widetilde\rho}^{\,(ee)}_{nm}
   -{\widetilde\rho}^{\,(ee)}_{nm}a^{\dag}a\right)
   \nonumber\\
   &+&\frac{1}{2}N\left(
   2a^{\dag}{\widetilde\rho}^{\,(ee)}_{nm}a
   -aa^{\dag}{\widetilde\rho}^{\,(ee)}_{nm}
   -{\widetilde\rho}^{\,(ee)}_{nm}aa^{\dag}\right)
   \nonumber\\
   &-&\frac{1}{2}M\left(
   2a^{\dag}{\widetilde\rho}^{\,(ee)}_{nm}a^{\dag}
   -\left(a^{\dag}\right)^{2}{\widetilde\rho}^{\,(ee)}_{nm}
   -{\widetilde\rho}^{\,(ee)}_{nm}\left(a^{\dag}\right)^{2}\right)
   \nonumber\\
   &-&\frac{1}{2}M^{*}\left(
   2a{\widetilde\rho}^{\,(ee)}_{nm}a
   -a^{2}{\widetilde\rho}^{\,(ee)}_{nm}
   -{\widetilde\rho}^{\,(ee)}_{nm}a^{2}\right)\,,
\end{eqnarray}
with
\begin{equation}
N\equiv\frac{k_{B}T}{\hbar\omega_{c}}-\frac{1}{2}\,,
\qquad
M\equiv-\frac{k_{B}T}{\hbar\omega_{c}}\,,
\end{equation}
and $\tau\equiv\gamma t$.
Eq.~(\ref{masee}) is a valid master equation, preserving the positivity of the density matrix
for $N(N+1)\ge|M|^{2}$ \cite{qnoise}
which occurs in the high temperature limit, i.e.
$k_{B}T\gg\hbar\omega_{c}$.

Since $\Theta_{nm}^{(ee)}$ represent normally ordered
characteristic functions, we can transform Eq.~(\ref{masee}) into
a partial differential equation \cite{qnoise}, that is
\begin{eqnarray}\label{Theq}
\frac{\partial}{\partial \tau}\Theta_{nm}^{(ee)}&=&\left\{
\left(i\frac{1+\chi}{\gamma}-\frac{1}{2}\right)
\lambda\frac{\partial}{\partial\lambda}\right.
\nonumber\\
&&\left.
-\left(i\frac{1+\chi}{\gamma}+\frac{1}{2}\right)\lambda^{*}
\frac{\partial}{\partial\lambda^{*}}
\right.\nonumber\\
&&\left.
+i\frac{\kappa}{\gamma}(n-m)\frac{\partial}{\partial\lambda}
-i\frac{\kappa}{\gamma}(n-m)\frac{\partial}{\partial\lambda^{*}}
\right.
\nonumber\\
&&\left.
-i\frac{\kappa}{\gamma} m\lambda-i\kappa n\lambda^{*}
\right.\nonumber\\
&&\left.
-\frac{1}{2}
\lambda\frac{\partial}{\partial\lambda^{*}}
-\frac{1}{2}
\lambda^{*}\frac{\partial}{\partial\lambda}
\right.\nonumber\\
&&\left.
-\frac{N}{2}
\lambda^{2}-\frac{N}{2}
\left(\lambda^{*}\right)^{2}
-N|\lambda|^{2}\right\}\Theta_{nm}^{(ee)}\,.
\nonumber\\
\end{eqnarray}
We then use a Gaussian trial function, which is justified by the choice of an initial Gaussian state
for the cantilever (a shifted thermal state) and by the fact that Eq.~(\ref{Theq}) preserves Gaussian states,
\begin{eqnarray}\label{trial}
\Theta_{nm}^{(ee)}(\tau)&=&\exp\left\{
A(\tau)\lambda+B(\tau)\lambda^{*}+C(\tau)|\lambda|^{2}
\right.\nonumber\\
&&\left.
D(\tau)\frac{\lambda^{2}}{2}+E(\tau)\frac{\left(\lambda^{*}\right)^{2}}{2}
+F(\tau)\right\}.
\end{eqnarray}
As a consequence, Eq.~(\ref{Theq}) yields the following
set of linear differential equations
\begin{eqnarray}
\label{Aeq}
\dot{A}(\tau)&=&\left(i\omega-\frac{1}{2}\right)A(\tau)
-\frac{1}{2}B(\tau)\nonumber\\
&&-i\frac{\kappa}{\gamma}(n-m)
\left(C(\tau)-D(\tau)\right)-i\frac{\kappa}{\gamma} m\,,
\\
\dot{B}(\tau)&=&-\frac{1}{2}A(\tau)-\left(i\omega+\frac{1}{2}\right)B(\tau)
\nonumber\\
&&+i\frac{\kappa}{\gamma}(n-m)\left(C(\tau)-E(\tau)\right)-i\frac{\kappa}{\gamma} n\,,
\\
\dot{C}(\tau)&=&-C(\tau)-\frac{1}{2}\left(E(\tau)+D(\tau)\right)-N\,,
\\
\dot{D}(\tau)&=&-C(\tau)+\left(2i\omega-1\right)D(\tau)-N\,,
\\
\dot{E}(\tau)&=&-C(\tau)-\left(2i\omega+1\right)E(\tau)-N\,,
\\
\label{Feq}
\dot{F}(\tau)&=&i\frac{\kappa}{\gamma}(n-m)\left(A(\tau)-B(\tau)\right)\,,
\end{eqnarray}
where we have set $\omega=(1+\chi)/\gamma$.
Accordingly to Eq.~(\ref{iniee}), initial conditions now read
\begin{eqnarray}
A(0)&=&\alpha^{*}\,,
\\
B(0)&=&-\alpha\,,
\\
C(0)&=&-\left(N+\frac{1}{2}\right)\,,
\\
D(0)&=&0\,,
\\
E(0)&=&0\,,
\\
F(0)&=&\ln\left[\frac{\beta^n\left(\beta^{*}\right)^{m}}
{\sqrt{n!m!}}e^{-|\beta|^{2}}\right]\,.
\end{eqnarray}
Notice that
\begin{equation}\label{Thetaee}
\Theta_{nm}^{\,(ee)}(\lambda=0,\tau)
={\rm Tr}_{c}\left\{{\widetilde\rho}^{\,(ee)}\right\}
=e^{F(\tau)}\,.
\end{equation}
Solving Eqs.~(\ref{Aeq})-(\ref{Feq}) we get
\begin{widetext}
\begin{eqnarray}\label{Ft}
F(\tau)&=&F(0)+\frac{i(\kappa/\gamma)(n-m)}{4\omega^{2}-16\omega^{4}}
\left\{-2\frac{F_{1}}{1+\Omega}\left[e^{-(1+\Omega)\tau/2}-1\right]
-2\frac{F_{2}}{1-\Omega}\left[e^{-(1-\Omega)\tau/2}-1\right]
\right.\nonumber\\
&&\left.
-\frac{F_{3}}{1+\Omega}\left[e^{-(1+\Omega)\tau}-1\right]
-\frac{F_{4}}{1-\Omega}\left[e^{-(1-\Omega)\tau}-1\right]
+F_{5}\left[e^{-\tau}-1\right]+F_{6}\tau\right\}\,,
\end{eqnarray}
where $\Omega^{2}=1-4\omega^{2}$ and
\begin{eqnarray}
F_{\tiny{\begin{array}{c}
1\\ 2\end{array}}}&=&
\frac{\kappa}{\gamma}(n-m)\left[
-4iN+16iN\omega^{2}\mp 8i\frac{\omega^{2}}{\Omega}\left(1+3N\right)
\pm32i\frac{\omega^{4}}{\Omega}\left(1+N\right)\pm
4i\frac{N}{\Omega}\right]
\nonumber\\
&&+\frac{\kappa}{\gamma}(n+m)\left[-2\omega+8\omega^{3}\pm 2\frac{\omega}{\Omega}
\mp 8\frac{\omega^{3}}{\Omega}\right]
\nonumber\\
&&+{\rm Re}\{\alpha\}\left[4\omega^{2}\pm 16\omega^{4}
+4\frac{\omega^{2}}{\Omega}\pm 16 \frac{\omega^{4}}{\Omega}\right]
+{\rm Im}\{\alpha\}\left[8\frac{\omega^{3}}{\Omega}
\pm 32\frac{\omega^{5}}{\Omega}\right]\,,
\\
F_{\tiny{\begin{array}{c}
3\\ 4 \end{array}}}&=&
\frac{\kappa}{\gamma}(n-m)\left[-2i+4i\omega^{2}
\pm 2i\frac{1}{\Omega}\mp 8i\frac{\omega^{2}}{\Omega}\right]\,,
\\
F_{5}&=&8i\omega^{2}\frac{\kappa}{\gamma}(n-m)\,,
\\
F_{6}&=&\frac{\kappa}{\gamma}(n-m)\left[4i+8iN-16i\omega^{2}-32i\omega^{2}N\right]
+\frac{\kappa}{\gamma}(n+m)\left[4\omega-16\omega^{3}\right]\,.
\label{F6}
\end{eqnarray}

\end{widetext}

\subsection{Solution for ${\tilde\rho}^{\,(gg)}$}

The quantities $\Theta_{nm}^{(gg)}(\lambda=0,\tau)$
are given by Eqs.(\ref{Thetaee})-(\ref{F6}) with the replacement $\chi\to -\chi$
(hence $(1-\chi)/\gamma\to\omega$).

\end{document}